\documentclass[12pt]{iopart}

\usepackage{iopams}
\usepackage{color}

\usepackage[utf8]{inputenc}       % Codepage
\usepackage[english]{babel} % Babel
\usepackage{dsfont}
\usepackage{bbm}
\usepackage[perpage,symbol*]{footmisc}
\usepackage[final]{graphicx}                             %   Graphics bundle
\usepackage[varg]{txfonts}
\begin{document}

\title[Post-Galilean  transformations of space and time derivatives and their consequences]{Post-Galilean  transformations of space and time derivatives and their consequences}

\author{Arbab I. Arbab and Faisal A. Yassein}

\address{Department of Physics, Faculty of Science, University of Khartoum, P.O. Box 321, Khartoum
11115, Sudan}
\ead{aiarbab@uofk.edu and f.a.yassein@gmail.com}

\begin{abstract}
Using post-Galilean space and time derivatives transformations and quantum mechanics, we have found  a new particle-wave equation besides the Klein-Gordon equation describing a spinless scalar particle. This new equation can also be obtained from Dirac's equation if $\beta=\gamma\left(1\pm\frac{v}{c}\right)$. Biot-Savart law and  additional continuity equations are obtained as a consequence of the invariance of Dirac's equation and Maxwell's equations under these transformations.
\end{abstract}

%Keywords: Electrodynamics, Fluid mechanics, Euler equation, Lorentz force, incompressible fluid, Mathematical formulation
%Uncomment for PACS numbers title message
% Keywords required only for MST, PB, PMB, PM, JOA, JOB?
%\vspace{2pc}
%\noindent{\it Keywords}: Article preparation, IOP journals
% Uncomment for Submitted to journal title message
%\submitto{\JPA}
% Comment out if separate title page not required
\maketitle
%\section{Introduction}
\section{introduction}
In this paper, we will use the post-Galilean transformation of space and time derivatives. Applying these transformations to equations describing the motion of a particle moving with constant velocity yields new physics. The space derivative transformation involves the Einstein's energy-momentum relation. The application of these derivatives transformations to Dirac's equation yields one of the continuity equations we have developed recently. The invariance of  Maxwell's equations to these transformations yields the Biot-Savart law.  Applying these transformations to the electric and magnetic fields shows that the electric and magnetic fields of a moving charged  particle are perpendicular to the velocity of the particle. The invariance of the physical equations under these derivative transformations yields other fundamental physical equations. Dirac's and Maxwell's equations are invariant under Lorentz transformations. However, the invariance of these equations under the post-Galilean transformations yields new physics.  The equations obtained above from the requirement of the invariance of post-Galilean transformations represent the physics of the particle moving at non-relativistic velocity. Any relativistic equations should reduce to these equations and not to the Newtonian ones.

Einstein's special theory of relativity hypothesized that physical laws must have the same form (covariant) for all inertial observers. Moreover, the speed of light in vacuum is constant irrespective of the velocity of the inertial frame in which it is measured. Accordingly, the relativistic mechanics is amended so that the mass, energy (kinetic $E_K$, and total $E$) and momentum ($p$) are related by [1]
\begin{equation}
E=E_K+m_0c^2\,, \qquad E^2=p^2c^2+m_0^2c^4\,, \qquad\vec{p}=\frac{\vec{v}}{c^2}\,E\,.
\end{equation}
Einstein equation of mass and energy is written in the form
\begin{equation}
 E^2=c^2p^2+E_0^2\,,\qquad E_0=m_0c^2\,.
\end{equation}
where the relativistic momentum $p$ is related to the non-relativistic momentum $p_{o}$ of the particle moving with velocity $\vec{v}$ by
\begin{equation}
 \vec{p}=\gamma \,\vec{p}_{o}\,,\qquad \gamma=\frac{1}{\sqrt{1-\frac{v^2}{c^2}}}\,.
\end{equation}
Moreover, one also has
\begin{equation}
 \vec{p}=\frac{\vec{v}}{c^2}E\,.
\end{equation}
In quantum mechanics, the  momentum and energy are defined by the operators [2]
\begin{equation}
 \vec{p}=-i\hbar\,\vec{\nabla}\,,\qquad E=i\hbar\frac{\partial}{\partial t}\,\,.
\end{equation}
Substituting Eq.(5) in Eq.(1) yields the Klein-Gordon equation for a scalar field $\varphi$, viz.,
\begin{equation}
\frac{1}{c^2}\frac{\partial^2\varphi}{\partial t^2}-\nabla^2\varphi+\left(\frac{m_0c}{\hbar}\right)^2\varphi=0\,,
\end{equation}
According to Eq.(5), Eq.(4) reads
\begin{equation}
 \vec{\nabla}=-\frac{\vec{v}}{c^2}\,\frac{\partial}{\partial t}\,\,.
\end{equation}
We define this equation as {\tt Einstein's energy - momentum operator equation}.
By squaring Eq.(7), one gets the operator equation
\begin{equation}
\frac{v^2}{c^2}\frac{1}{c^2}\frac{\partial^2}{\partial t^2}-\nabla^2=0\,,
\end{equation}
which when acts on a function $\psi$ yields the wave equation,
\begin{equation}
\frac{v^2}{c^2}\frac{1}{c^2}\frac{\partial^2\psi}{\partial t^2}-\nabla^2\psi=0\,.
\end{equation}
For a photon, one has
\begin{equation}
\frac{1}{c^2}\frac{\partial^2\psi}{\partial t^2}-\nabla^2\psi=0\,,
\end{equation}
where $v=c$. This is the familiar wave equation for a massless particle. Notice here that Eq.(9) is another equation that describes the scalar particle that is equivalent to the Klein-Gordon equation. However, unlike the Klein-Gordon equation, Eq.(9) doesn't embody the mass of the article  explicitly. For a plane wave solution, i.e., $\psi\propto \exp -i(\omega t-k x)$, Eq.(9) yields
\begin{equation}
\frac{c^2}{v}=\frac{\omega}{k}=v_p\,,
\end{equation}
where $v_p$ is the phase velocity of the wave. Thus, Eq.(9) represents the  wave equation of the familiar de Broglie's wave. Equation (11) gives the familiar relation $v=\frac{c^2}{v_p}$ between the particle and wave velocity in the wave-packet interpretation of quantum mechanics. Accordingly, Eq.(9) can be written as
\begin{equation}
\frac{1}{v_p^2}\frac{\partial^2\psi}{\partial t^2}-\nabla^2\psi=0\,.
\end{equation}
This is a wave equation valid for massive and massless particles. But for massless  particles $v_p=c$. The mass of the particle is hidden in the particle velocity $v$, since $\frac{v}{c}=\sqrt{1-\left(\frac{m_0}{m}\right)^2}$. This is unlike the Klein-Gordon equation where the mass of the particle appears explicitly in the equation.
\section{Rest energy operator}
In quantizing the Einstein energy equation the rest mass energy is not described by an appropriate operator. In this section we will find out this operator. To this end, we apply Eqs.(4) in Eq.(1) to obtain
\begin{equation}
E_0^2=E^2-c^2p^2=\left(1-\frac{v^2}{c^2}\right)E^2
\Rightarrow\qquad E_0=\sqrt{1-\frac{v^2}{c^2}}\,\,E=\frac{E}{\gamma}\,.
\end{equation}
Hence, using Eqs.(2) and (4), the rest mass operator is
\begin{equation}
E_0=i\hbar\,\frac{\partial}{\gamma\,\partial t}\,.
\end{equation}
This equation suggests that the rest mass operator becomes
\begin{equation}
\hat{m}_0=\frac{i\hbar}{c^2\gamma}\,\frac{\partial}{\partial t}\,.
\end{equation}
This amounts to say that the mass is a dynamical quantity and not just some scalar value. This equation compliments Eq.(9) and provides the mass of the particle described by the wave function $\psi$. It is interesting  that applying Eq.(15) to Eq.(1) yields Eq.(8).
\section{Post-Galilean transformations}
We call the non-relativistic Lorentz transformation
 \begin{equation}
t=t'+\frac{v}{c^2}x\,'\,,\qquad x=x\,'+v\,t\,'\,,\qquad y=y\,'\,,\qquad z=z\,'\,.
\end{equation}
the post-Galilean transformations. The space and time partial derivatives are then
\begin{equation}
\frac{\partial}{\partial t\,'}=\frac{\partial}{\partial t}+\vec{v}\cdot\vec{\nabla}\,,
\end{equation}
and
\begin{equation}
\vec{\nabla}\,'=\vec{\nabla}+\frac{\vec{v}}{c^2}\frac{\partial}{\partial t}\,.
\end{equation}
The continuity equation  relating the probability current density, $\vec{J}$ and  probability density, $\rho$
\begin{equation}
\vec{\nabla}\cdot \vec{J}+\frac{\partial\rho}{\partial t}=0\,,
\end{equation}
under these transformation yields
\begin{equation}
\vec{\nabla}\cdot \vec{J}+\frac{\partial\rho}{\partial t}=0\,,\qquad \vec{\nabla}\rho+\frac{1}{c^2}\frac{\partial \vec{J}}{\partial  t}=0\,.
\end{equation}
These two equation have been shown to be part of a generalized continuity equations [5]
\begin{equation}
\vec{\nabla}\cdot \vec{J}+\frac{\partial\rho}{\partial t}=0\,,\qquad \vec{\nabla}\rho+\frac{1}{c^2}\frac{\partial \vec{J}}{\partial  t}=0\,,\qquad \vec{\nabla}\times\vec{J}=0\,.
\end{equation}
These equations are invariant under the transformations in Eqs.(17) and (18).
Bear in mind that the three above equations in Eq.(21) are Lorentz invariant. These equations can be written as
\begin{equation}
\partial_\mu J^\mu=0\,,\qquad\partial_\mu J_\nu-\partial_\nu J_\mu=0\,,\qquad J^\mu=\left(\rho c\,, \vec{J}\right)\,.
\end{equation}
If we consider the current and density transform under post-Galilean as
\begin{equation}
 \vec{J}\,'=\vec{J}-\rho\,\vec{v}\,,\qquad \rho\,'=\rho-\frac{\vec{v}\cdot\vec{J}}{c^2}\,.
\end{equation}
In this case the continuity equation will be invariant under the transformations in Eqs.(17) and (18).
It interesting to notice that Klein-Gordon equation is invariant under the transformations in Eqs.(17) and (18).
\section{Dirac's equation}
Dirac's equation can be written as
\begin{equation}
\frac{\partial \psi}{\partial t}=-\,c\,\vec{\alpha}\cdot\vec{\nabla}\psi-\frac{i\beta\,m_0c^2}{\hbar}\,.
\end{equation}
When this equation is subject to the transformations in Eqs.(17) and (18), one obtains
\begin{equation}
\frac{\partial \psi}{\partial t}+\vec{v}\cdot\vec{\nabla}\psi=-\vec{\alpha}\cdot\frac{\vec{v}}{c}\frac{\partial \psi}{\partial t}-\,c\,\vec{\alpha}\cdot\vec{\nabla}\psi-\frac{i\beta\,m_0c^2}{\hbar}\,,
\end{equation}
where we have assumed that $\psi\,'(x')=\psi(x)$.
Applying Eq.(24) in Eq.(25), yields
\begin{equation}
\vec{v}\cdot\vec{\nabla}\psi=-\vec{\alpha}\cdot\frac{\vec{v}}{c}\frac{\partial \psi}{\partial t}\,.
\end{equation}
Take the adjoint of Eq.(26), and multiply it by $\psi$ once and multiply Eq.(26) by $\psi^+$ from right. The addition of the two resulting equations yields
\begin{equation}
\vec{v}\cdot\left(\vec{\nabla}\rho+\frac{1}{c^2}\frac{\partial \vec{J}}{\partial t}\right)=0\,,\qquad{\rm where} \qquad \rho=\psi^+\psi,\qquad \vec{J}=\psi^+(c\vec{\alpha})\psi\,,
\end{equation}
where $\vec{\alpha}^+=\vec{\alpha}$ \, and \,$\beta^+=\beta$. Equation (27) is true if
\begin{equation}
\vec{\nabla}\rho+\frac{1}{c^2}\frac{\partial \vec{J}}{\partial t}=0\,.
\end{equation}
This equation is nothing but one of our continuity equations in Eq.(21). Thus, Dirac's equation is compatible with the generalized continuity equations in Eq.(21), as we have already shown [5].
\section{Electromagnetism and Einstein's operator equation}
The electric and magnetic fields  are given by
\begin{equation}
\vec{E}=-\vec{\nabla}\varphi-\frac{\partial \vec{A}}{\partial t}\,,\qquad \vec{B}=\vec{\nabla}\times\vec{A}\,.
\end{equation}
The Lorentz gauge connecting the two fields is given by
\begin{equation}
\vec{\nabla}\cdot\vec{A}+\frac{1}{c^2}\frac{\partial\varphi }{\partial t}=0\,.
\end{equation}
When this equation is subject to the transformations in Eqs.(17) and (18), one obtains
\begin{equation}
(\vec{\nabla}\cdot\vec{A}+\frac{1}{c^2}\frac{\partial\varphi }{\partial t})+
\vec{v}\cdot\,(\vec{\nabla}\varphi+\frac{\partial \vec{A}}{\partial t})=0\,.
\end{equation}
Applying Eqs.(29) and (30), one obtains the condition
\begin{equation}
\vec{v}\,\cdot\vec{E}=0\,.
\end{equation}
Hence, the Lorentz gauge is invariant under Eqs.(17) and (18), assuming the fields don't change, if the electric field lines  are perpendicular to the particle's velocity.
Now apply Eqs.(17) and (18)  to the electric field in Eq.(29) to get
\begin{equation}
(-\vec{\nabla}\varphi-\frac{\partial \vec{A}}{\partial t})+\vec{v}\,(\vec{\nabla}\cdot\vec{A}+\frac{1}{c^2}\frac{\partial\varphi }{\partial t})=0\,.
\end{equation}
This clearly shows that the electric field is invariant under the transformation defined by Eqs.(17) and (18), provided that the Lorentz gauge is satisfied.
If we now assume the potentials $\vec{A}$ and $\varphi$ do change under post-Galilean  transformation, then
\begin{equation}
\vec{A}\,'=\vec{A}-\frac{\vec{v}}{c^2}\,\varphi\,,\qquad \varphi\,'=\varphi-\vec{v}\cdot\vec{A}\,.
\end{equation}
Applying Eqs.(34), (17) and (18) to  Eq.(29) yields
\begin{equation}
\vec{B}\,'=\vec{\nabla}\,'\times\vec{A}\,'=\vec{\nabla}\times\vec{A}+\frac{\vec{v}\times\vec{E}_q}{c^2}=\vec{B}+\vec{B}_q\,,
\end{equation}
and
\begin{equation}
\vec{E}\,'=E-\vec{v}\times\vec{B}_q\,,
\end{equation}
The magnetic field $\vec{B}_q=\frac{\vec{v}\times\vec{E}_q}{c^2}$ and the electric fields $\vec{E}_q=-\vec{v}\times\vec{B}_q$represent the magnetic and electric fields of a charge moving with constant velocity. This is given by the Biot-Savart law.

We now apply the transformations in Eqs.(17), (18),  (35) and (36) to Maxwell's equations
\begin{equation}
\vec{\nabla}\times\vec{E}=-\frac{\partial\vec{B} }{\partial  t}\,.
\end{equation}
\begin{equation}
\vec{\nabla}\times\vec{B}=\mu_0\vec{J}+\frac{1}{c^2}\frac{\partial\vec{E}}{\partial t}\,,
\end{equation}
and
\begin{equation}
\vec{\nabla}\cdot\vec{B}=0\,,\qquad \vec{\nabla}\cdot\vec{E}=\frac{\rho}{\varepsilon_0}\,.
\end{equation}
The invariance of Eq.(37) to Eqs.(17) and (18) requires
\begin{equation}
\vec{B}_q=\frac{\vec{v}}{c^2}\times\vec{E}_q\,,
\end{equation}
so that
\begin{equation}
\vec{\nabla}\times\vec{E}_q=-\frac{\partial\vec{B}_q }{\partial  t}\,,
\end{equation}
which shows that the $\vec{B}_g$ and $\vec{E}_g$ satisfy Faraday's equation too.
The  invariance of Eqs.(39) under Eqs.(17) and (18) imply that
\begin{equation}
\vec{v}\cdot\vec{B}_q=0\,,\qquad \vec{v}\cdot\vec{E}_q=0\,.
\end{equation}
The invariance of Eq.(38) under the transformations Eqs.(17), (18), (35) and (36) implies that
\begin{equation}
\vec{\nabla}\times\vec{B}_q=\mu_0\vec{J}+\frac{1}{c^2}\frac{\partial }{\partial  t}(-\vec{v}\times \vec{B}_q)\,,\qquad \vec{J}=\rho\,\vec{v}\,.
\end{equation}
We infer from this equation that the electric field of a moving particle is
\begin{equation}
\vec{E}_q=-\vec{v}\times \vec{B}_q\,.
\end{equation}
This is also evident from Eq.(36).
Hence, the electric and magnetic fields of a moving charge satisfy Maxwell's equations [3, 6].

The Lorentz force on a charged particle is given by
\begin{equation}
\vec{F}=q(\vec{E}+\vec{v}\times \vec{B})\,.
\end{equation}
Applying the transformations in Eqs.(17) and (18) and using  Eqs.(35), (36) and (44) yield
\begin{equation}
\vec{F}\,'=q(\vec{E}\,'+\vec{v}\times \vec{B}\,')=q(\vec{E}+\vec{v}\times \vec{B})\,.
\end{equation}
Hence, the Lorentz force is invariant under the post-Galilean transformations.
\section{concluding remarks}
We have developed in this paper two post-Galilean transformation of space and time derivatives with which the particle's dynamics can be expressed. Invariance of Dirac's equation and  Maxwell's equations yield the dynamic and the generalized continuity equation that we already developed. The equations obtained above from the requirement of the invariance of post-Galilean transformations represent the physics of the particle moving at non-relativistic velocity. Any relativistic equations should reduce to these equations and not to the Newtonian ones.
\section*{References}
$[1]$ Rindler, W., \emph{Introduction to Special Relativity}, Oxford University Press, USA (1991).\\
$[2]$ Griffiths, D., \emph{Introduction to elementary particles}, John Wiley, (1987).\\
$[3]$ Arbab, A. I. and Yassein, F. A., \emph{A new formulation of quantum mechanics}, {\tt submitted to Chinese physics letters}, 2009.\\
$[4]$ Arbab, A. I. and Yassein, F. A., \emph{A new formulation of electrodynamics}, {\tt submitted to Chinese physics letters}, 2009.\\
$[5]$ Arbab, A. I. and Widatallah, H.M., \emph{The generalized continuity equation}, {\tt submitted to Chinese physics letters}, 2009.\\
$[6]$ Arbab, A. I. and Satti, Z. A., \emph{On the generalized Maxwell equations and their prediction of electroscalar wave}, \emph{Progress in Physics}, \textbf{2}, 8 (2009).\\

\end{document}